# Measurement of the Pressure induced by salt crystallization in confinement

*J. Desarnaud, D. Bonn, N. Shahidzadeh.*

Van der Waals-Zeeman Institute, Institute of Physics, University of Amsterdam, Science Park 904, 1098 XH Amsterdam, The Netherlands

Salt crystallization is a major cause of weathering of artworks, monuments and rocks. Damage will occur if crystals continue to grow in confinement, i.e. within the pore space of these materials generating mechanical stresses. We report on a novel method that allows to directly measure, at the microscale, the resulting pressure while visualizing the spontaneous nucleation and growth of alkali halide salts. The experiments reveal the crucial role of the wetting films between the growing crystal and the confining walls for the development of the pressure. The results suggest that the pressure originates from a charge repulsion between the similarly charged wall and the crystal separated by a ~1.5 nm salt solution film. Consequently, if the walls are made hydrophobic, no film and no crystallization pressure are detected. The magnitude of the pressure is system-specific and explains how a growing crystal exerts stresses at the scale of individual grains in porous materials.

**TOC GRAPHICS**

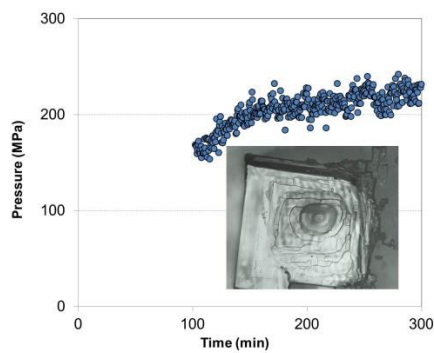

**KEYWORDS**-Crystallization, confinement, pressure, surface charges, damage, porous media

Salt weathering affects monuments, engineering structures, artworks, rock outcrops and minerals within the soil profile. There is compelling evidence that its influence will increase due to the global climate change[1]. Many ancient structures and historical artworks, such as the Valley of the Kings (Egypt), the Petra monument (Jordania), and a large number of frescos and sculptures have already been partially or even completely destroyed by salt attack[2-10] (Figure.1).

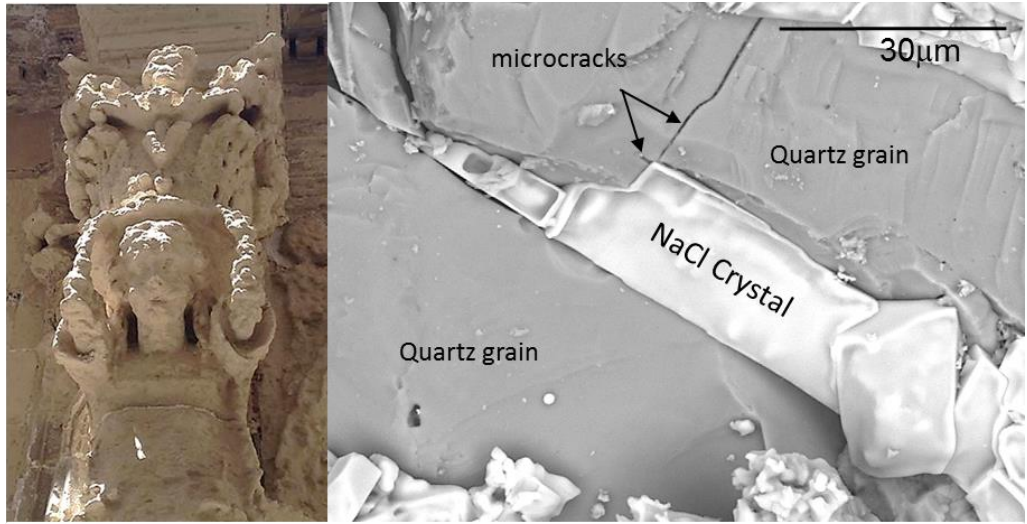

**Figure. 1**. **Left**: Degradation of a historical sandstone sculpture (Lecce, Italy). **Right**: SEM image of NaCl crystals precipitation (white) in the pore space of sandstone after evaporation of salt solution. Two perpendicular mirocracks can be observed where the NaCl crystal has formed.

Arguments explaining that growing crystals can generate stress have been given for more than 150 years ago, since Lavalle and subsequently Becker reported that a crystal can grow under external pressure[11,12]. They assumed that the crystal could grow and overcome the external forces provided the surface is in contact with supersaturated solution[13-16]. In its simplest form, the crystallization pressure $P_{crystallization}$ was approximated by the chemical potential difference between salt molecules in the crystal and in the supersaturated salt solution: $P_{crystallization} \sim \nu RT/V_m \ln(S)$ with $\nu$ the number of ions, with R the gas constant, T the temperature, $V_m$ the molar volume of the solid phase, and S the supersaturation, which is the ratio of the solute concentrations in the solution $c$ and in the saturated solution $c_s$.[14-17]. Although several theoretical corrections have been made on this equation by taking into account the water activities and the size of the crystal[18-20], it does not fully explain the different experimental results reported in literature, notably the very different weathering effect of different salts. The direct measurement of the force exerted by a growing crystal in confinement is challenging, as illustrated by the small number of experimental results reported[14,21-23]. Previous studies consider a pre-existing crystal immersed in a salt solution of a given salinity with the crystal confined between two plates under a load. They reported the growth kinetics of the confined face[23], the displacement of the wall due to the growth [13,14,21] or the repulsive force between a crystal face and a wall using AFM[22]. The debate about the mechanism involved in the development of such a pressure continues, especially since thermodynamically a pressure occurs only in presence of a supersaturation. More formally, this requires an activity product K[P] greater than the solubility product at atmospheric pressure K[0], as described by Freundlich equation for a crystal under pressure $P = \frac{RT}{V} \ln \frac{K[P]}{K[0]}$. In addition, the crystal could simply grow in a direction in which it is not confined[15,21,25].

We report experimental results obtained with an innovative setup which permits to visualize the spontaneous nucleation and growth during the evaporation of the salt solution while directly measuring the force exerted by a crystal during its growth in confinement.

Two alkali halides salts, sodium chloride (NaCl) and potassium chloride (KCl), are studied firstly because of their abundance and similar crystalline stucture (i.e. rock-salt structure). The second incentive for this choice of materials is that recently in experiments on the evaporation of NaCl solutions it became clear that primary nucleation results in very high supersaturations S =$c/c_s$~1.6±0.2 at the onset of crystallization[26]. This should be contrasted with earlier attempts to measure the pressure exerted during a secondary nucleation process (growth of pre-existing crystal), for which very modest supersaturations were considered[21]. Here, we study the spontaneous primary nucleation and growth of one crystal and do not impose a certain supersaturation, or impose a given load; we rather measure the force that the growing crystal exerts with its spontaneous precipitation. As the exerted pressure scales with ln(S), these recent findings suggest a scenario in which the force exerted by a growing crystal should be easier detected than in the previous experiments.

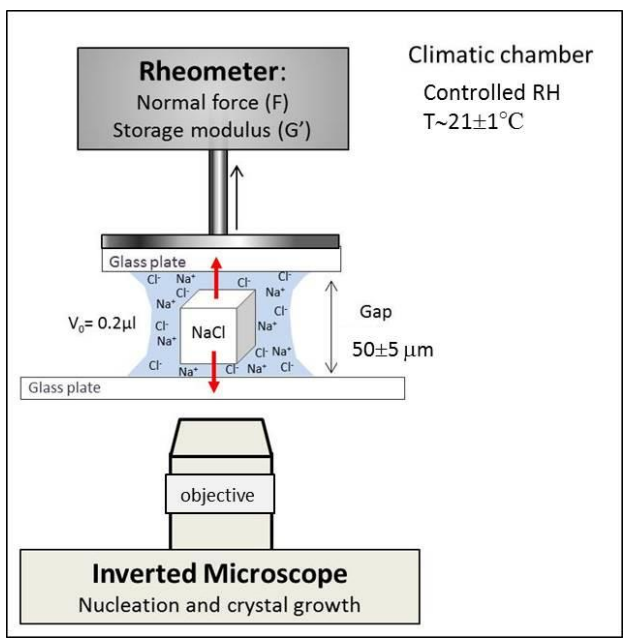

**Scheme 1**: Setup for the visualization of the spontaneous nucleation and growth of a crystal in microscale confinement and the simultaneous detection of the force exerted by the growing crystal. The spontaneous precipitation of the microcrystal is induced by controlled evaporation of the salt solution at temperature T = 21±1°C at relative humidities RH = 40±3% or RH=4±2%.

Our experimental setup is as follows (Scheme 1). A very small drop ($V_0 \sim 0.2$ μl) of the salt solution (NaCl or KCl) close to saturation (S = c/c$_s$ = 0.9, with $c_s^{NaCl} = 6.16$ mol.$kg^{-1}$, $c_s^{KCl} = 4.56$ mol.$kg^{-1}$) is deposited on a cleaned glass slide on an inverted microscope. The second glass slide that is used to confine the drop from above with a microscale gap is attached to a mechanical testing machine (a rheometer) that allows to (i) control and measure the gap between the plates, (ii) detect the normal force on the plate during the crystal growth, and (iii) deduce the surface area of the salt crystal that is in contact with the plates. The latter is done by applying very small oscillations to the upper plate and measuring the resistance of the material between the two plates to a small shear force By direct visualization using a CCD camera coupled to the inverted optical microscope, we have followed during the evaporation, the volume change of the entrapped salt solution, the onset of the crystal precipitation, and the crystal growth in the solution. (see Supplementary Information).

Once the salt solution is entrapped between the two glass plates, an attractive (negative) capillary force (F~-3.5 10$^{-3}$ ± 1.0 10$^{-3}$ N) is detected; the measured force is in good agreement with the calculated theoretical value by considering the gap distance, the surface tension of the salt solution and its contact angle with the glass plate[27] (see Supplementary Information). Upon evaporation of the water, generally a single crystal is observed to form. In the first 5 seconds after its nucleation, the crystal size grows to fill the gap (50 μm) with a film of solution between the crystal faces and the glass plates (i.e. the crystal is observed to be able to move) (Figure 2).

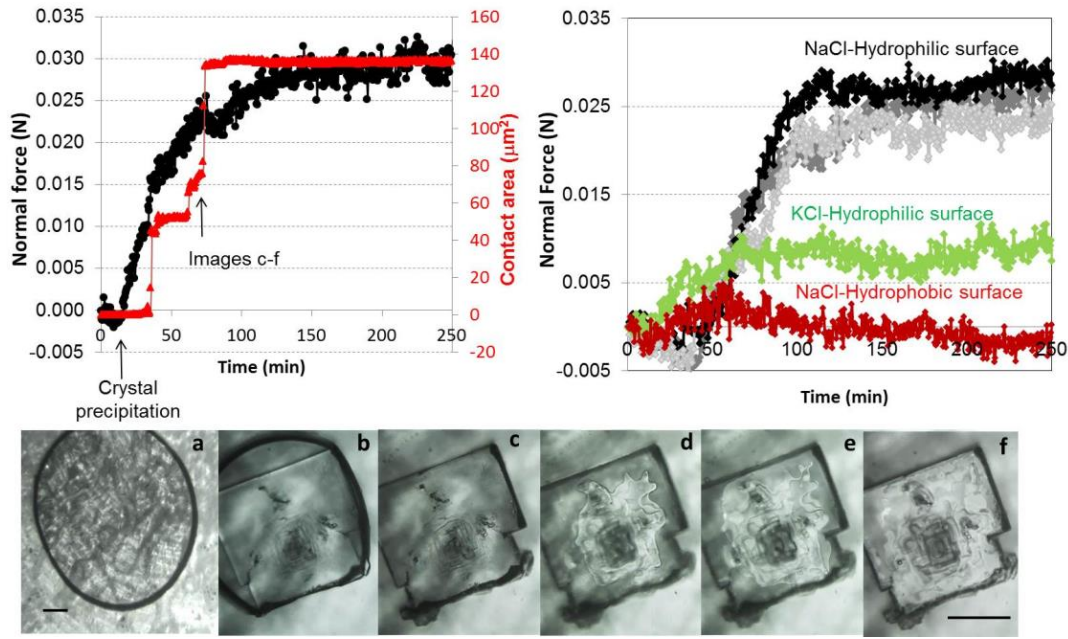

**Figure 2**: Top Left: Evolution of the normal force and the crystal-glass contact area with the NaCl growth from the solution during evaporation. Top Right: (**Black curves**): 3 experiments of NaCl growth between hydrophilic glass slides showing the reproducibility; measured Force for a KCl crystal between two hydrophilic glass plates (**green curve**) and on NaCl crystal between two hydrophobic glass plates (**red curve**). Bottom: (a) evaporation of the NaCl solution entrapped between two hydrophilic glass plates before crystal precipitation ; (b-f) Crystal precipitation and growth and the drying of the liquid film between the crystal and the glass plates. The variation of the grey level in images e and f is due to the evolution of steps in the confined face of the crystal. Development of efflorescence is visible in the left corner of the main crystal (image c-f); Scale bars: 100 μm.

With the further growth of the crystal, we find that a repulsive (positive) force develops, increasing with increasing the evaporation time. The maximum force is achieved when there is no more solution around the sides of the crystal (i.e. no more lateral growth): the remaining salt solution is in the liquid film between the glass plates and the crystal. Its subsequent evaporation is observed to induce the formation of steps on the confined face of the crystal[23] and the precipitation of small microcrystallites ('efflorescence') [28,29] around the main crystal (Figure 2 & 3). For NaCl crystals, the maximum value of the force $\Delta F \sim 0.03 \pm 0.07$ N is achieved when the remaining film of salt solution is thinner than the optical resolution of the microscope (Figure 2).

To identify the origin of the crystallization pressure, we first convert the measured force into a pressure by measuring the area of contact using the rheometer (see supplementary information). The latter gives the shear modulus of any solid material that occupies the gap between the two plates, by measuring the shear stress necessary for a certain imposed deformation. The bulk salt solution is liquid and does not resist shear, whereas the salt crystal does[30]. The mechanical behaviour of very thin films is less clear; consequently, for the calculation of the pressure we use the apparent shear modulus at the end of drying where no more water is left. Separate measurements at ~ 4% relative humidity show comparable results at the end of drying. The growth of the crystal is indeed observed to lead to a strong increase of the apparent shear modulus that increases in a similar fashion as the force, although more intermittently. This allows to obtain the filling degree and hence the real contact area at the end of the experiment (~136 $\mu m^2$ for the experiment of Figure 2). For this experiment, the crystallization pressure deduced from the measured force and contact area is about 230±50 MPa, a very large pressure that is sufficient to damage e.g., the sandstone of Figure 1. This pressure is comparable to the yield stress of NaCl[31] and larger than earlier measurement of disjoining pressure[22,32]. However, the latter were performed on different salts with a different type of glass and/or with mica, so that the surface charges maybe different. At the end of experiments, the height of the NaCl microcrystals was measured using SEM. Surprisingly, this is found to be around 120±20 μm (Figure 3), i.e. larger than the fixed gap between the two plates (~50 µm) which explains why a plateau is reached for the measured force at the ned of drying. This is possible since the bottom plate is a thin microscope slide that bends during the experiments, as is observed directly by the changed focus of the microscope objective during the growth of the crystal. These results obtained at the microscale confirm once again the arguments given at the beginning of century that a growing crystal in confinement can lift a load if the latter is in contact with a film of solution[12,13].

The importance of thin liquid films for the crystallization pressure was already underlined theoretically[11-20]: in order for a crystallization pressure to develop, one needs to maintain a film of liquid by some repulsive forces between the crystal and the wall as this allows for continued crystal growth by the addition of extra ions to the crystal lattice. Once a growing crystal fully bridges the gap between two confining walls, no crystallization pressure can be developed since no extra layers of salt can be added to give rise to such a pressure. To demonstrate this, we have performed the same experiment with glass plates that are made hydrophobic by a silanization treatment. Indeed, in this case, as the formation of a wetting film is prevented, no repulsive force is detected during growth within the experimental uncertainty (Figure 2). This in good agreement with the very low crystallization pressure measured for the growth of NaCl crystal in contact with hydrophobic PDMS miro channels[24].

These experimental results make DLVO-type thin-film forces an obvious candidate for the origin of the crystallization pressure[18,33]. The Van der Waals forces through a thin film of salt solution between a salt crystal and a glass surface are attractive, and thus could not lead to a pressure pushing the two surfaces apart. A charge repulsion between the two surfaces, provided they have the same charge. If this were the case, the *disjoining pressure* $\Pi(l)$ within a film of thickness *l,* pushing the two surfaces apart, would be of the (simplified) form $\Pi(l) \sim A \exp(-\kappa l)$, with $A = \frac{2\sigma^2}{\varepsilon\varepsilon_0}$, σ the charge density at the surface, ε and $\varepsilon_0$ the relative permittivity of the bulk and free space, respectively, and $\kappa^{-1} = \frac{0,304}{\sqrt{[NaCl]}}$ (nm) the Debye screening length with [NaCl] the molar sodium chloride concentration[33].

To compare our results with the expression for Π (l), we need to determine the NaCl concentration in the solution during the establishment of the pressure. The supersaturation in the solution at the onset of crystallization is the one we can determine with the greatest accuracy, the error in these measurement is about ΔS= ±0.05; After precipitation, the supersaturation does not change very rapidly. Salt crystals grows only logarithmically slowly due to the slowness of the diffusion of ions $D_{Na+} \sim 1.3 \ 10^{-9}$ m$^2$/s[34,35]; the evaporation is fast compared to the crystal growth. Also due to the slowness of diffusion, over a typical time scale in the experiment (~1000s) there will be little exchange between the entrapped film and the remaining solution besides the crystal; the typical size of the reservoir is ~1 cm, whereas the diffusion length is at most a fraction of a mm. The experimental observation that the increase of the force and surface area are similar also suggest that during drying the pressure remains almost constant, suggesting in turn that the supersaturation in the entrapped salt solution between the confined crystal face does not change.

To check that the salt concentration in the liquid film between the crystal and the wall remains as high as the concentration at the onset of crystallization, we stop the experiment when the maximum force and maximum shear modulus are achieved but a thin liquid film is still visible under the microscope (t ~ 150 min). The crystal is gently removed and after further drying the upper crystal face (i.e. confined face) is observed using scanning-electron microscopy (SEM). The microphotographs reveal the formation of large amount of tiny dentritic crystals on the crystal face (Figure 3) that form with the evaporation of the thin layer of concentrated solution when the crystal is removed from the setup.

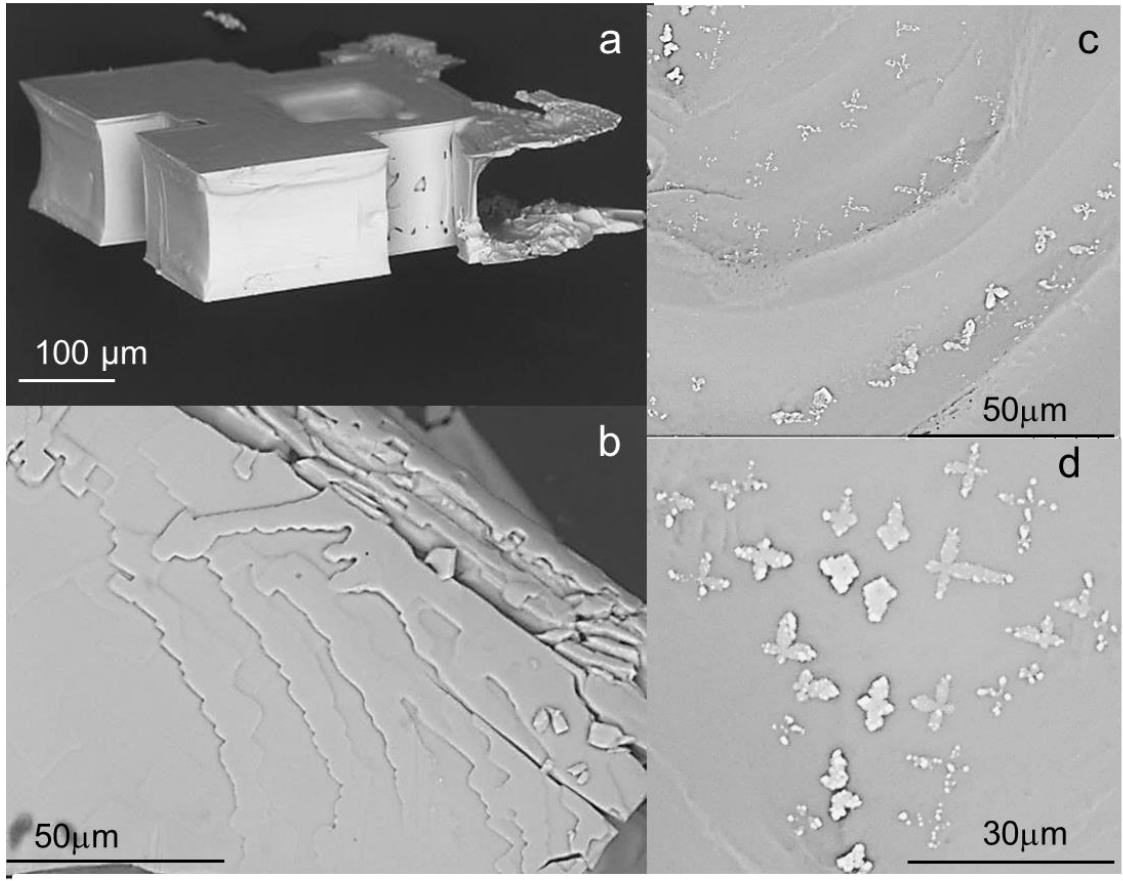

**Figure 3:** SEM images of NaCl crystals : (a) the profile view and (b) the confined face of the crystal at the end of the experiment ( t=300min); (c&d) The confined face of NaCl crystal half-way through drying (t=150 min) ; precipitation of large amount of tiny dentritic crystals caused by the evaporation of the highly concentrated film of salt solution.

Consequently, if we assume that the concentration in the film equals the supersaturation at the onset of crystallization, the measured pressure as a function of the supersaturation follows an exponential decay, in excellent agreement with expression for the disjoining pressure, with a constant film thickness $l$ of about 1.5 nm (Figure 4a); this value is again in very good agreement with the expected magnitude for such a film ($l \sim 2$ nm)[17,32,33]. Moreover, these results are also in agreement with a similar repulsive interaction reported between a silica-covered AFM tip and a potassium sulfate crystal separated by a film of salt solution[22]. It is perhaps surprising that this simple DLVO-type expression [33], gives such a good description of our data; both non linear terms in the Poisson-Boltzman treatment of this problem could become important at these high saturations, as well as non-DLVO forces such as hydration forces [22,32,37-39]. This merits further discussion, but is beyond the scope of this paper.

The scenario of a repulsive force also requires that the crystal/solution and the solution/glass interfaces carry a charge of identical sign. Although it is known that glass surfaces are charged negatively when in contact with water and a low concentration of salt[40], it is less evident that this remains the case if they are in contact with a *highly* saturated salt solution[41-43]. Consequently, we have determined the surface charges in highly concentrated salt solutions (i.e. close to saturation) by measuring the adsorption of cationic ((Methylene Blue / $C_{16}H_{18}ClN_3S$, at $8.10^{-6}$ M)) and anionic dye (Eosine Y / $C_{20}H_6Br_4Na_2O_5$, at $1.2.10^{-5}$ M) onto glass and crystals surfaces in contact with a highly concentrated salt solution. Methylene blue adsorption is widely used for the dermination of both surface areas and exchange capacities of clay minerals [44]. Using UV/Vis-Spectrometer one can measure the de-coloration of a dye solution due to the adsorption of the dye on the oppositely charged surfaces (in our case on glass surface and salt crystals), and determine in this way the nature of the charge of a given surface and its density (see Supplementary Information). These experiments show that glass beads, NaCl and KCl are all negatively charged since they adsorb the cationic dye. This therefore indeed allows for the development of a repulsive disjoining pressure (Figure 4).

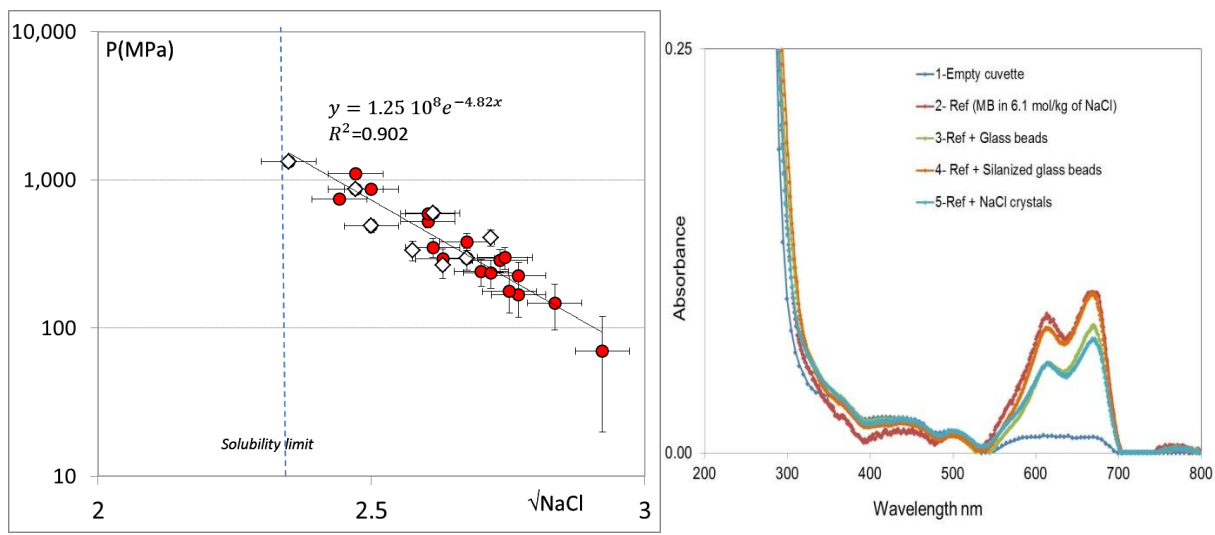

**Figure 4: Left:** The final measured pressure induced by the primary nucleation and growth of NaCl between two hydrophilic glass plates as a function of the square root of the salt concentration at the onset of crystallization. White square symbols: experiments at RH ~4%; red circles : at RH~40%. The solid line is a fit to the expression for the repulsive force between two charged surfaces. **(b)** UV-Vis Spectrometer experiments for 1) an empty cuvette 2) the reference solution (Ref): The Methylene Blue (MB) in 6.1 mol.kg$^{-1}$ of NaCl solution 3,4,5) The referencesolution in contact with glass beads, silanised glass beads and 2g of NaCl crystals added to the solution.

More quantitatively, this allows to estimate the surface charge density $\sigma$ at the silica surface in contact with water and both NaCl and KCl solutions. In pure water we find $\sigma \sim 0.22$ C.m$^{-2}$ in good agreement with reported values[33] *(see supplementary information)*. The surface charge density of glass decreases in contact with salt solutions: $\sigma \sim 0.12$ C.m$^{-2}$ for NaCl and $\sigma \sim 0.072$ C.m$^{-2}$ for KCl. The lower charge density for the KCl solution is probably due to the larger size of K$^+$ (ionic radius of 1.33 Å) compared to Na+ (0.96 Å), which results in smaller hydrated radii. Due to a smaller radius, the K$^+$ ions bind more easily to the negatively charged silica surface at high concentrations[44]. Since the surface charge features directly in the amplitude A of the disjoining pressure, one would anticipate a much smaller crystallization pressure for KCl compared to NaCl. Indeed, if we repeat the experiments of force measurements with KCl solution, we find much smaller pressures; in a typical experiment P ~ 70 ±15 MPa - roughly an order of magnitude smaller than for NaCl in also with the significantly smaller surface charge of KCl (Figure 2). Reflecting these lower crystallization pressures we also find much less damage when we expose pieces of sandstone to multiple dissolution/drying cycles using a KCl solution compared to NaCl[45]. Unfortunately, the range of supersaturations achieved in the several experiments with KCl was very narrow (S = 1.25 ± 0.05), which does not allow us to determine the disjoining pressure curve for this case.

Thus, the dependence of the fore on salt concentration, surface charge and wettability of the glass makes a strong point that the observed forces are due to the pressure of thin liquid film between charges surfaces. It is interesting to compare the thermodynamic crystallization pressure[19], which is the maximum pressure that the wall will exert in order to avoid the growth of the crystal at a given supersaturation, and the disjoining pressure opposing the contact between the crystal and the wall. Our results show that for supersaturations higher than 1.4, the disjoining pressure and the crystallization pressure are very similar. For example, for a supersaturation of S ~ 1.5, the calculated thermodynamic pressure is on the order of 135 MPa [19], similar to the disjoining pressure measured here (150 MPa±50). This is an important finding, not only because the measured pressure is so high, but also because it is in excess of the tensile strength of the sandstone (~ 0.9 MPa[46]). With the crystallization pressure exceeding the tensile strength, crystallization can thus induce damage in the stone.

In conclusion, we have performed a direct measurement of the pressure exerted by the spontaneous precipitation of a salt microcrystal during evaporation in a confinement similar to that found in porous materials The results obtained for the growth of KCl and NaCl crystals between hydrophilic and hydrophobic glass walls suggests that the pressure originates from a repulsive interaction between charged surfaces separated by the liquid film. The magnitude of the measured pressure can be very high, i.e. in excess of the tensile strength of various stones used in artworks, sculptures and monuments and may account for the damage due to crystal growth on porous materials. The subtle dependence of the generated stress on the wetting and the surface charge properties that we report can also explain the experimentally observed variability for different salts crystallizing in different materials[14,21-24]. The understanding of processes that are at the origin of alterations of porous artworks is of great importance for managing and preserving them for future generations: hydrophobic treatments have for instance been successfully used in the past[38]; here we show experimentally how such treatments could prevent damage[16].

**Supporting information Available**

The supporting information includes the Experimental procedure, the determination of the attractive capillary force, the crystal-wall measurement of the contact area and the surface charges for glass, NaCl and KCl crystals in salt solutions.

ACKNOWLEDGEMENT

We thank the two anonymous referees for their very useful comments.